\documentclass[aps, reprint, twocolumn, superscriptaddress, showkeys]{revtex4-2}

\usepackage[utf8]{inputenc}
\usepackage[T1]{fontenc}
\usepackage[a4paper, margin=2cm]{geometry}
\usepackage[english]{babel}
\usepackage{textgreek}
\usepackage{dcolumn}
\usepackage{amsmath, amssymb}
\usepackage[separate-uncertainty=true, multi-part-units=single]{siunitx}
\usepackage{graphicx}
\usepackage{booktabs}
\usepackage{setspace}
\usepackage{natbib}
\graphicspath{{img/}}

\DeclareUnicodeCharacter{2212}{\ensuremath{-}}

\makeatletter
\renewcommand{\p@subsection}{}
\renewcommand{\p@subsubsection}{}
\makeatother

\begin{document}

\title{Effects of Disorder on the Magnetic Properties of the Heusler Alloy V\textsubscript{2}FeAl}
\author{R. Smith}
\email[Correspondence: ]{rsmith1@tcd.ie}
\affiliation{School of Physics, The University of Dublin Trinity College, College Green, Dublin 2, D02 K8N4, Ireland}
\author{Z. Gercsi}
\affiliation{School of Physics, The University of Dublin Trinity College, College Green, Dublin 2, D02 K8N4, Ireland}
\author{R. Zhang}
\affiliation{School of Physics, The University of Dublin Trinity College, College Green, Dublin 2, D02 K8N4, Ireland}
\affiliation{Northwest Institute for Nonferrous Metal Research, Xi'An 710016, China}
\author{K.E. Siewierska}
\affiliation{Helmholtz-Zentrum Berlin für Materialien und Energie, Germany}
\author{K. Rode} 
\affiliation{School of Physics, The University of Dublin Trinity College, College Green, Dublin 2, D02 K8N4, Ireland}
\author{J.M.D. Coey}
\affiliation{School of Physics, The University of Dublin Trinity College, College Green, Dublin 2, D02 K8N4, Ireland}
\date{\today}

\begin{abstract}
	Magnetic properties of multicomponent alloys depend sensitively on the degree of atomic order on the different crystallographic sites.
	In this work we demonstrate the magnetic contrast between bulk and thin-film samples of the Heusler alloy V\textsubscript{2}FeAl.
	Arc-melted bulk ingots show practically no site preference of the elements (A2 structure), whereas magnetron-sputtered thin-film samples display a higher degree of atomic ordering with a tendency towards XA-type order.
	Electronic structure calculations favour ferrimagnetic XA-type ordering, and the effect of different pairwise atomic disorder on the element specific and net magnetic moments are evaluated to reproduce experimental observations.
	XA-type thin-films with iron moment of 1.24 \textmu\textsubscript{B} determined by X-ray magnetic circular dichroism are in agreement with calculation, but the measured net moment of 1.0 \textmu\textsubscript{B} per formula unit and average vanadium moment are smaller than expected from calculations.
	The measured Curie temperature is approximately \SI{500}{\kelvin}.
	Films with a higher degree of disorder have a T\textsubscript{C} close to \SI{300}{\kelvin} with a net moment of 0.1 \textmu\textsubscript{B} at low temperature.
	The large calculated vanadium moments are destroyed by partial disorder on $4d$ vanadium sites.
	By contrast, the arc-melted and annealed bulk alloy with a fully-disordered A2 structure shows no spontaneous magnetization; it is a Pauli paramagnet with dimensionless susceptibility \textchi\textsubscript{v} = \SI{-2.95e-4}{}.
\end{abstract}

\maketitle

\section{Introduction}

\begin{figure}[!ht]
	\centering
	\includegraphics[width=0.5\linewidth]{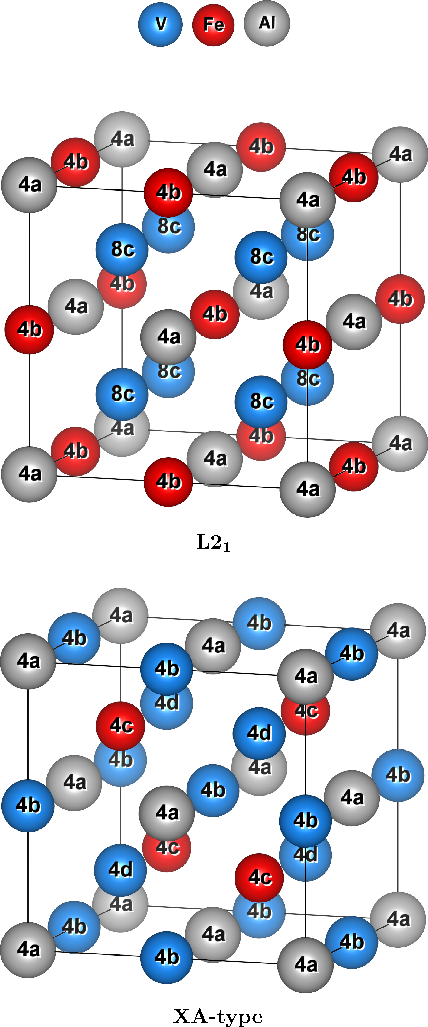}
	\caption{The two possible fully-ordered crystal structures of V\textsubscript{2}FeAl; L2\textsubscript{1} {(top)} where the vanadium, iron, and aluminium atoms sit in the $8c$, $4b$, and $4a$ Wyckoff sites respectively, and XA-type {(bottom)} where the iron and aluminium occupy the $4c$ and $4a$ sites respectively and the vanadium atoms sit in $4b$ and $4d$ sites.}\label{fig:V2FeAl_structures}
\end{figure}

Heusler alloys are materials with formula X\textsubscript{2}YZ (where X and Y are transition metals and Z is a p-block element), which consist of four interpenetrating face-centered cubic lattices~\cite{Graf2011}.
The Heusler family is vast, with thousands of possible combinations, and over 800 papers are published on the topic annually.
The alloys exhibit a wide variety of mechanical, electronic, and magnetic properties, and have applications in many disparate areas of condensed matter physics ranging from spintronics~\cite{Inomata2008}~\cite{Palmstrom2016} to thermoelectric power generation~\cite{Hu2020} and shape-memory behaviour~\cite{Chernenko2005}.
The growing global demand for cobalt has driven a recent focus on cobalt-free magnetic Heusler alloys, which aside from economic and environmental concerns, also offer advantages for certain spintronics applications, such as reduced magnetisation and high coercive fields with the goal of achieving faster dynamics for current-induced magnetization switching for example. 
The magnetic and transport properties of these materials are highly dependent upon the crystal structure and symmetry, as well as the level of atomic disorder. 
Here we contrast the structural and magnetic properties of V\textsubscript{2}FeAl in bulk and thin-film form; full Heusler alloys can adopt one of two possible fully-ordered crystal structures, L2\textsubscript{1} and XA (or inverse)-type, presented in Figure~\ref{fig:V2FeAl_structures}, as well as partially-ordered vacanies depending the processing conditions.

In the L2\textsubscript{1}-type crystal structure all of the vanadium atoms occupy the crystallographically equivalent $8c$ Wyckoff sites, with the iron and aluminium occupying the $4b$ and $4a$ sites respectively. 
The L2\textsubscript{1} structure has space group Fm$\bar{3}$m {(No. 225)}, with corresponding centrosymmetric point group symmetry m$\bar{3}$m.
Alternatively, in the XA-type structure, the vanadium atoms occupy two crystallographically inequivalent Wyckoff positions, $4b$ and $4d$, the iron now occupying the $4c$ sites and aluminium remaining in the $4a$ sites. 
This structure has space group F$\bar{4}$3m {(No. 216)}, with corresponding non-centrosymmetric point group symmetry $\bar{4}$3m.
As the vanadium atoms in the XA-type crystal structure occupy two crystallographically inequivalent sites, they form two inequivalent magnetic sublattices.
These are the fully-ordered crystal structures of V\textsubscript{2}FeAl, but there are also possible fully or partially disordered structures listed below and summarised in Table~\ref{tab:V2FeAl_structures}; 
\begin{enumerate}
	\item B32 structure in which there is partial disorder between the aluminium atoms in the $4a$ sites and the vanadium atoms in the $4b$ and $4d$ sites.
	\item DO\textsubscript{3} structure in which there is partial disorder between the iron atoms in the $4c$ sites and the vanadium atoms in the $4b$ and $4d$ sites.
	\item B2 structure in which there is partial disorder between the iron in the $4c$ sites and the aluminium in the $4a$ sites.
	\item A2 structure is  where atoms in all atomic sites are fully disordered. 
\end{enumerate}

\begin{table}[!ht]
	\centering
	\caption{The possible ordered and disordered structures of V\textsubscript{2}FeAl where $\leftrightarrow$ represents disorder between elements in a given site, space group number presented in parentheses.}\label{tab:V2FeAl_structures}
	\begin{tabular}{ccc}
		\toprule
		Struct. & Space Group & Disorder\\
		\midrule
		XA & F$\bar{4}$3m (216) & Fully-Ordered\\
		L2\textsubscript{1} & Fm$\bar{3}$m (225) & Fully-Ordered\\
		DO\textsubscript{3} & Fm$\bar{3}$m (225) & V\textsuperscript{$4d$} $\leftrightarrow$ Fe\textsuperscript{$4c$}\\
		B32 & Fd$\bar{3}$m (227) & V\textsuperscript{$4b$\&$4d$} $\leftrightarrow$ Al\textsuperscript{$4a$}\\
		B2 & Pm$\bar{3}$m (221) & Fe\textsuperscript{$4c$} $\leftrightarrow$ Al\textsuperscript{$4a$}\\
		A2 & Im$\bar{3}$m (229) & V\textsuperscript{$4b$\&$4d$} $\leftrightarrow$ Fe\textsuperscript{$4c$} $\leftrightarrow$ Al\textsuperscript{$4a$}\\
		\bottomrule
	\end{tabular}
\end{table}

V\textsubscript{2}FeAl has not been reported experimentally either in bulk or thin-film form, in contrast to a number of published electronic structure calculations.  
Watson et al.~\cite{Watson1998} place V\textsubscript{2}FeAl in the L2\textsubscript{1} structure and reported an unrealistically large unit cell parameter of \SI{1103.3}{\pico\meter} with ferrimagnetic ordering.
They concluded that the system is unstable due to a low heat of formation of -0.02 eV.
Kumar et al.~\cite{Kumar2009} also placed V\textsubscript{2}FeAl in the L2\textsubscript{1} structure with a more realistic lattice parameter and found that ferrimagnetic ordering is preferred.
Zhang et al.~\cite{Zhang2012} and Skaftouros et al.~\cite{Skaftouros2013} both determined the XA-type structure to be energetically preferred to the L2\textsubscript{1} structure, and predicted ferrimagnetism to be the preferred magnetic ordering, a result  corroborated by density functional theory (DFT) calculations.
The published electronic structure calculation results are summarised in Table~\ref{tab:V2FeAl_DFT}.

\begin{table*}[!ht]
	\centering
	\caption{Results of density functional theory calculations from different works with the preferred structure, magnetic ordering, cubic lattice parameter (pm), element specific moments (\textmu\textsubscript{B}), and total moment (\textmu\textsubscript{B} per unit cell). * Element specific moments add up to 0.77 \textmu\textsubscript{B}, not 0.46 \textmu\textsubscript{B}.}\label{tab:V2FeAl_DFT}
	\begin{tabular}{ccccccccc}
		\toprule
		Reference & Structure & Magnetic-Ordering & Lattice Constant & \textmu\textsubscript{Fe} & \multicolumn{2}{c}{\textmu\textsubscript{V}} & \textmu\textsubscript{Al} & \textmu\textsubscript{Tot.}\\
		\midrule
		Watson et al.~\cite{Watson1998} & L2\textsubscript{1} & FiM & 1103.4 & 1.15 & \multicolumn{2}{c}{-0.19} & 0.00 &  0.77\\	
		Kumar et al.*~\cite{Kumar2009} & L2\textsubscript{1} & FiM & 597.8 & 1.91 & \multicolumn{2}{c}{-0.57} & 0.00 & 0.46\\	
		Skaftouros et al.~\cite{Skaftouros2013} & XA & FiM & 593.0 & 1.20 & 2.11 & -0.31 & -0.09 & 2.92\\
		Zhang et al.~\cite{Zhang2012} & XA & FiM & 592.0 & 1.18 & 2.46 & -0.46 & -0.18  & 3.00\\
		\bottomrule
	\end{tabular}
\end{table*}

\section{Experimental}

Epitaxial thin-films of V\textsubscript{2}FeAl were grown by DC magnetron sputtering on $10\mathrm{mm}\times10\mathrm{mm}$ single-crystal MgO{(001)} substrates in the ultra-high vacuum Trifolium Dubium sputtering system with a base pressure less than \SI{1e-9}{} mbar.
The films were co-sputtered from high-purity \SI{50}{\milli\meter} targets of vanadium and binary iron-aluminium {(50:50)} with a confocal sputtering geometry.
Films were grown at a range of deposition temperatures from \SIrange{300}{700}{\degreeCelsius}.
The MgO substrates were degassed for \SI{1}{\hour} at \SI{600}{\degreeCelsius} to reduce surface contamination and encourage epitaxial film growth.
Deposition rates were calculated based upon the thicknesses and densities of V and FeAl calibration samples measured by X-ray reflectivity.
Following sample deposition, the thin-films were transferred under ultra-high vacuum to the dedicated \textit{in-situ} X-ray photoelectron spectroscopy chamber equipped with a Specs PHOIBOS 150 hemispherical energy analyser and monochromated Al K\textsubscript{\textalpha1} X-ray source.
The analyser was set to transmission mode with an acceptance area determined by the X-ray spot size (\SI{2}{\milli\meter}); all spectra were captured with a pass energy of 10 eV, step size of 0.1 eV, and dwell time of \SI{1}{\second}.
Fitting of XPS spectra was carried out using the CasaXPS software package in order to confirm the chemical composition of the films and check for contamination. 
A fluorescent RHEED screen was used to probe surface crystallinity.
A \SI{2}{\nano\meter} capping layer of Al\textsubscript{2}O\textsubscript{3} was then deposited by radio-frequency magnetron sputtering to protect the sample against oxidation.
Bulk ingots were prepared by arc-melting high-purity elemental samples of vanadium, iron, and aluminium followed by grinding the ingot into a powder and annealing at temperatures ranging from \SIrange{600}{900}{\degreeCelsius}.

A Panalytical X'Pert Pro X-ray diffractometer using Cu K\textsubscript{\textalpha1} radiation (\textlambda\ = \SI{0.15406}{\nano\meter}) was used to capture diffraction patterns of both bulk powder and thin-film samples.
Low-angle X-ray reflectivity was also measured on the thin-films, and the open source GenX XRR refinement programme~\cite{Bjorck2007} was used to fit the measured data and determine sample density, thickness, and roughness.
Roughness was confirmed by atomic force microscopy {(Veeco Nanoscope III atomic force microscope with Multimode software suite)}.
Reciprocal space mapping of the thin-film samples were performed on a Bruker D8 Discover high-resolution X-ray diffractometer with a Cu K\textsubscript{\textalpha1} beam from a double-bounce Ge{(220)} monochromator.
Rietveld refinement of the powder diffraction patterns was carried out using FullProf~\cite{Rodriguez1990}.

Magnetization measurements were performed with a \SI{5}{\tesla} Quantum Design MPMS-XL SQUID magnetometer.
A \SI{57.2}{\milli\gram} sample of bulk powder was measured, as were thin-films with the field applied perpendicular to the film plane.
The diamagnetic contribution from the MgO substrate was corrected for including a paramagnetic contribution from Mn\textsuperscript{2+}/Fe\textsuperscript{3+} impurities in the MgO which appears below about \SI{100}{\kelvin}~\cite{Venkatesan2014}.
A model consisting of a Curie-Weiss law paramagnetic component {($m_{Para}$)}, a diamagnetic component {($m_{Dia}$)}, and a spontaneous magnetic moment arising from the V\textsubscript{2}FeAl thin-film {($m_{Sample}$)} was used to fit magnetization curves at \SI{5}{\kelvin}, \SI{100}{\kelvin}, and \SI{300}{\kelvin} as well as temperature scans in \SI{2}{\tesla} to ensure the sample was saturated.
The paramagnetism in the samples is modelled using a Brillouin function;

\begin{equation}
	\begin{gathered}
		m_{Para}=ng\mu_{B}J\bigg[\frac{2J+1}{2J}\coth\bigg(\frac{2J+1}{2J}\cdot x\bigg)\\
		+\frac{1}{2J}\coth\bigg(\frac{x}{2J}\bigg)\bigg],\\
		\mathrm{where}\,\,\, x=\frac{g\mu_{B}J}{k_{B}(T-\theta_{P})}\cdot\mu_{0}H_{Applied},
	\end{gathered}
\end{equation}

with $J=\frac{5}{2}$. 
We then refine the number of paramagnetic ions ($n$), and the paramagnetic Curie temperature ($\theta_{P}$), to account for magnetic interactions between the ions (the magnitude of this effect is typically less than \SI{1}{\kelvin}). 
The diamagnetic substrate contribution is simply;
			
\begin{equation}
	m_{Dia}=\chi_{V}\cdot H_{Applied},
\end{equation}

where $\chi_{V}\le0$ is the diamagnetic susceptibility which we refine, ensuring the value doesn't differ appreciably from values cited in literature. 
Finally, the temperature dependence of the magnetizatiom is modelled using Bloch's T\textsuperscript{3/2} law;
			
\begin{equation}
	m_{Ferro}=m_{0}\bigg[1-{\bigg(\frac{T}{T_{C}}\bigg)}^{3/2}\bigg],
\end{equation}
	
where we fit the spontaneous magnetization at zero temperature ($m_{0}$), and the Curie temperature ($T_{C}$). 
Note that 3\textsigma\ noise removal has been performed on the magnetometry data.

Iron L-edge X-ray absorption near-edge structure {(XANES)} and X-ray magnetic dichroism {(XMCD)} of V\textsubscript{2}FeAl thin-films were measured at the VEKMAG beamline of the BESSY II light source of Helmholtz-Zentrum Berlin.
Measurements were performed on two samples, one ordered and the other disordered.
The XANES of the Fe-L\textsubscript{III} and Fe-L\textsubscript{II} edges  were measured over an energy range of 680 eV to 780 eV in an applied field of \SI{2}{\tesla}.
The absorption is measured in total electron yield, whereby a drain current from the sampleis measured, and normalised to a mirror current from the final optical component along the beampath before the sample (this mirror shouldn't show any absorption and the mirror current should therefore be directly proportional to the intensity of incident X-rays).
The absorption spectra are measured for both left circularly polarised (LCP) and right circularly polarised (RCP) X-rays, then again with the applied field direction reversed as the spectrum measured with RCP X-rays and positive applied field should be equivalent to that measured with LCP and negative applied field.
The sum of these spectra gives the XANES spectrum, while the difference gives the XMCD spectrum.
A background function is subtracted from the XANES spectra prior to analysis.
This function consists of a component linear in energy with a pair of arctangent functions at the L\textsubscript{III} and L\textsubscript{II} edges, the amplitude of the latter being half that of the former.
The integrated areas of the XANES and XMCD spectra are then calculated and the spin and orbital moments deduced using the XMCD sum rules~\cite{Chen1995}.
The moments must be scaled to the number of holes in the 3d band.
In this work we chose a value of n\textsubscript{h} = 3.3, based upon electronic structure calculations of the XA-type structure for the ordered thin-film sample.
The number of holes for the disordered sample was then determined by comparing the magnitude of the edge-jumps against the ordered sample and then scaling the value of n\textsubscript{h} accordingly.
The number of holes in the disordered sample was determined to be approximately half that of the ordered sample.

Thin-film samples were patterened photolithographically into Hall bars with length and width of \SI{50}{\micro\meter} and \SI{10}{\micro\meter} respectively.
Ru/Ta/Pt contact pads were fabricated by a lift-off technique, which were subsequently cold-welded to \SI{50}{\micro\meter} diameter silver wire using high purity indium.
Anomalous Hall effect measurements were carried out using a \SI{5.5}{\tesla} superconducting magnet with a cryostat capable of reaching temperatures down to \SI{10}{\kelvin}.
Transport measurements were performed with a DC current of \SI{0.5}{\milli\ampere}.

Ab-initio calculations based on density functional theory were carried out using norm-conserving pseudopotentials and pseudo-atomic localized basis functions implemented in the OpenMX software package~\cite{Ozaki2004}.
The generalized gradient approximation {(GGA-PBE)}~\cite{Perdew1996} was used for all calculations.
We used a 16 atom convenient cell for the cubic structure using $15\times15\times15$ k-points to evaluate the total energies of the ordered and site-disordered structures.
Pre-generated fully relativistic pseudopotentials and the pseudo-atomic orbitals with a typical cut-off radius of 7 atomic units (a.u.) with s3p3d3 were used for all elements, respectively.
An energy cut-off of 300 Ry was used for the numerical integrations.
The convergence criterion for the energy minimization procedure was set to 10\textsuperscript{-7} Hartree.
Spin-orbit interaction (SOI) was turned off and only collinear spin configurations were considered.

\section{Results}

\begin{figure*}[!ht]
	\centering
	\includegraphics[width=\textwidth]{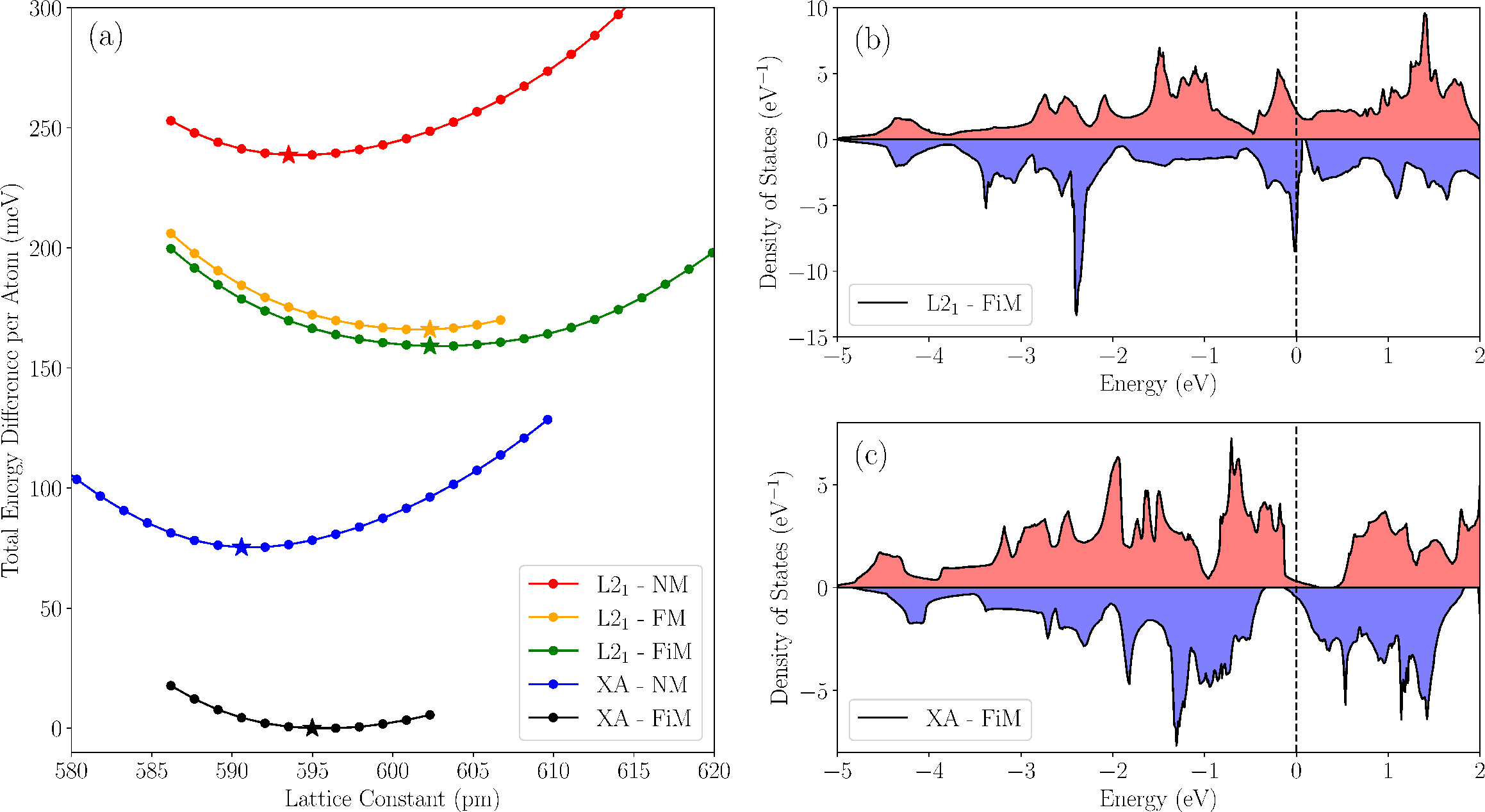}
	\caption{(a) Total energy difference among the different structures and magnetic orderings of V\textsubscript{2}FeAl as a function of lattice parameter calculated by density functional theory, the star markers represent the lattice constant corresponding to the lowest total energy (b) Density of states for L2\textsubscript{1} structured V\textsubscript{2}FeAl with ferrimagnetic ordering, (c) Density of states for XA-type structured V\textsubscript{2}FeAl with ferrimagnetic ordering.}\label{fig:V2FeAl_DFT}
\end{figure*}

\begin{table}[!ht]
	\centering
	\caption{Results of density functional theory calculations for the possible fully-ordered structures and magnetic-orderings of V\textsubscript{2}FeAl, showing the cubic lattice parameters (pm), element specific moments (\textmu\textsubscript{B}), and total moments (\textmu\textsubscript{B} per formula unit).}\label{tab:V2FeAl_DFT_Zsolt}
	\begin{tabular}{cccccccc}
		\toprule
		Struct. & Ord. & a & \textmu\textsubscript{Fe} & \multicolumn{2}{c}{\textmu\textsubscript{V}} & \textmu\textsubscript{Al} & \textmu\textsubscript{Tot.}\\
		\midrule
		L2\textsubscript{1} & NM & 593.5 & --- & \multicolumn{2}{c}{---} & --- & ---\\	
		L2\textsubscript{1} & FM & 602.3 & 1.90 & \multicolumn{2}{c}{0.20} & 0.00 & 2.30\\
		L2\textsubscript{1} & FiM & 602.3 & 2.10 & \multicolumn{2}{c}{-0.86} & -0.07 & 0.31\\	
		XA & NM & 590.6 & --- & --- & --- & --- & ---\\
		XA & FiM & 595.0 & 1.26 & 1.88 & -0.29 & 0.05 & 2.90\\
		\bottomrule
	\end{tabular}
\end{table}

The results of density functional theory calculations performed for different crystal structures and magnetic-orderings of V\textsubscript{2}FeAl are shown in Table~\ref{tab:V2FeAl_DFT_Zsolt}.
Figure~\ref{fig:V2FeAl_DFT} (a) shows that the XA-type structure is energetically favourable compared to the L2\textsubscript{1} structure, and that ferrimagnetic (FiM) ordering is preferred over non-magnetic or ferromagnetic ordering in both cases.
The calculation of the XA-type structure with ferromagnetic ordering did not converge suggesting a highly unfavourable energetic spin configuration.
The results for FiM XA-type V\textsubscript{2}FeAl agree with the previous reports by Skaftouros et al.~\cite{Skaftouros2013} and Zhang et al.~\cite{Zhang2012}, only with a slightly larger cubic lattice parameter of \SI{595.0}{\pico\meter}, a slightly higher moment on the iron atoms, and lower moment on the vanadium.
The L2\textsubscript{1} results predict a larger lattice parameter than Kumar et al.~\cite{Kumar2009}, with larger element specific moments, but a lower net moment.
In this structure, the iron atoms in the $4b$ sites and take on a large moment of 2.10 \textmu\textsubscript{B} which is mostly cancelled by the antiferromagnetically coupled vanadium moments (-0.86 \textmu\textsubscript{B}) in the $8c$ sites, giving a small net moment of 0.31 \textmu\textsubscript{B} per unit cell.
In the XA-type structure, the Fe atoms occupy the $4c$ site previously taken by V, showing a smaller moment of 1.26 \textmu\textsubscript{B}.
The vanadium atoms remaining in the $4d$ sites take on a large moment of 1.88 \textmu\textsubscript{B} coupled ferromagnetically to the Fe moment with the remaining vanadium in the $4b$ sites having a small negative moment of -0.29 \textmu\textsubscript{B}.
This gives the XA-type structure a total moment of 2.90 \textmu\textsubscript{B} per formula unit, which matches the moment of 3 \textmu\textsubscript{B} predicted by the Slater-Pauling rule {(V\textsubscript{2}FeAl has 21 valence electrons $\Rightarrow$ $M=24-Z=3$ \textmu\textsubscript{B})}.

\begin{table*}[!ht]
	\centering
	\caption{Effects of disorder on the magnetic moments of V\textsubscript{2}FeAl, element specific moments are in \textmu\textsubscript{B} per atom, total moment is in \textmu\textsubscript{B} per formula unit.}\label{tab:V2FeAl_DFT_Disorder}
	\begin{tabular}{cccccccccc}
		\toprule
		Structural & \multicolumn{2}{c}{\textmu\textsubscript{Fe}} & & \multicolumn{2}{c}{\textmu\textsubscript{V\textsubscript{I}}} & & \textmu\textsubscript{V\textsubscript{II}} & \textmu\textsubscript{Al} & \textmu\textsubscript{Tot.}\\
		Ordering & ($4b$ & $4c$) & \textmu\textsubscript{Fe}\textsuperscript{avg} & ($4b$ & $4c$) & \textmu\textsubscript{V\textsubscript{I}}\textsuperscript{avg} & $(4d$) & ($4a$) & f.u. \\
		\midrule
		100\% XA & --- & 1.26 & 1.26 & -0.29 & --- & -0.29 & 1.88  & 0.05 & 2.90 \\
		75\% XA / 25\% L2\textsubscript{1} & 1.98 & 1.64 & 1.73 &  -0.35 & -0.51 & -0.39 & 1.47 & 0.00 & 2.81 \\ 
		50\% XA / 50\% L2\textsubscript{1} & 2.02 & 1.87 & 1.95 & -0.26 & -0.46 & -0.36 & 0.99 & -0.02 & 2.56 \\ 
		25\% XA / 75\% L2\textsubscript{1} & 2.16 & 0.20 & 1.67 & 0.16 & -0.49 & -0.33 & -1.07 & -0.07 & 0.20 \\ 
		100\% L2\textsubscript{1} & 2.10 & --- & 2.10 & --- & -0.86 & -0.86 & -0.86 & -0.07 & 0.31 \\
		\bottomrule
	\end{tabular}
\end{table*}

The effects of binary disorder on the magnetic properties of V\textsubscript{2}FeAl have also been explored.
Table~\ref{tab:V2FeAl_DFT_Disorder} shows the site and element specific moments as one goes from the XA-type structure toward the L2\textsubscript{1} structure {(i.e.\ starting from the fully-ordered XA-type structure, one sequentially swaps iron atoms in the $4c$ sites with vanadium atoms in the $4b$ sites until the fully-ordered L2\textsubscript{1} structure is reached.)}
It is apparent that as disorder is introduced into the system, the magnetization decreases relative to the two fully-ordered structures.
A drastic change is observed in the net moment between 50\% XA and 25\% XA, largely due to the vanadium atoms in the $4d$ sites now coupling antiferromagnetically to the iron atoms in the $4b$ and $4c$ sites.
Note that in the XA-type structure, the vandium atoms in the $4d$ sites take on a large moment of 1.88 \textmu\textsubscript{B} coupled ferromagnetically to the iron in the $4c$ sites.
This moment is much greater than has been previously reported for vanadium in metallic systems. 
For example, spin-polarised electronic structure calculations on binary FeV reveal a moment of -0.29 \textmu\textsubscript{B}~\cite{Jordan1991}, which is consistent with what we observe for vanadium atoms in $4b$ and $4c$ sites.
Polarised neutron scattering studies on bulk bcc Fe-V binary systems have shown that vanadium can have a moment up to approximately -3 \textmu\textsubscript{B}, but this is only observed for very dilute concentrations of vanadium {(\textless\ 1 at.\%)}~\cite{Kajzar1980}.
At higher vanadium concentrations the moment decreases dramatically and at 10 at.\% a moment of approximately -1 \textmu\textsubscript{B} is observed~\cite{Mirebeau1987}, which then falls monotonically with increasing vanadium concentration, with the alloys becoming non-magnetic at 77 at\% vanadium~\cite{Lam1963}.
V\textsubscript{2}FeAl contain 50\% vanadium, so one might expect a moment of about 0.4 \textmu\textsubscript{B} or less.
We therefore believe that the calculated moment of the vanadium in the $4d$ sites (1.88 \textmu\textsubscript{B} in fully-ordered XA structure) to be overestimated in partially-disordered thin-films, and that in reality they have a moment closer in magnitude to those of the $4b$ and $4c$ sites (0.29 \textmu\textsubscript{B}) or to what is observed in the L2\textsubscript{1} structure (0.86 \textmu\textsubscript{B}).
Making this assumption, we find the total moment of the XA-type structure falls in the range from 1.31 \textmu\textsubscript{B} to 1.88 \textmu\textsubscript{B} per formula unit.

\begin{figure*}[!ht]
	\centering
	\includegraphics[width=\textwidth]{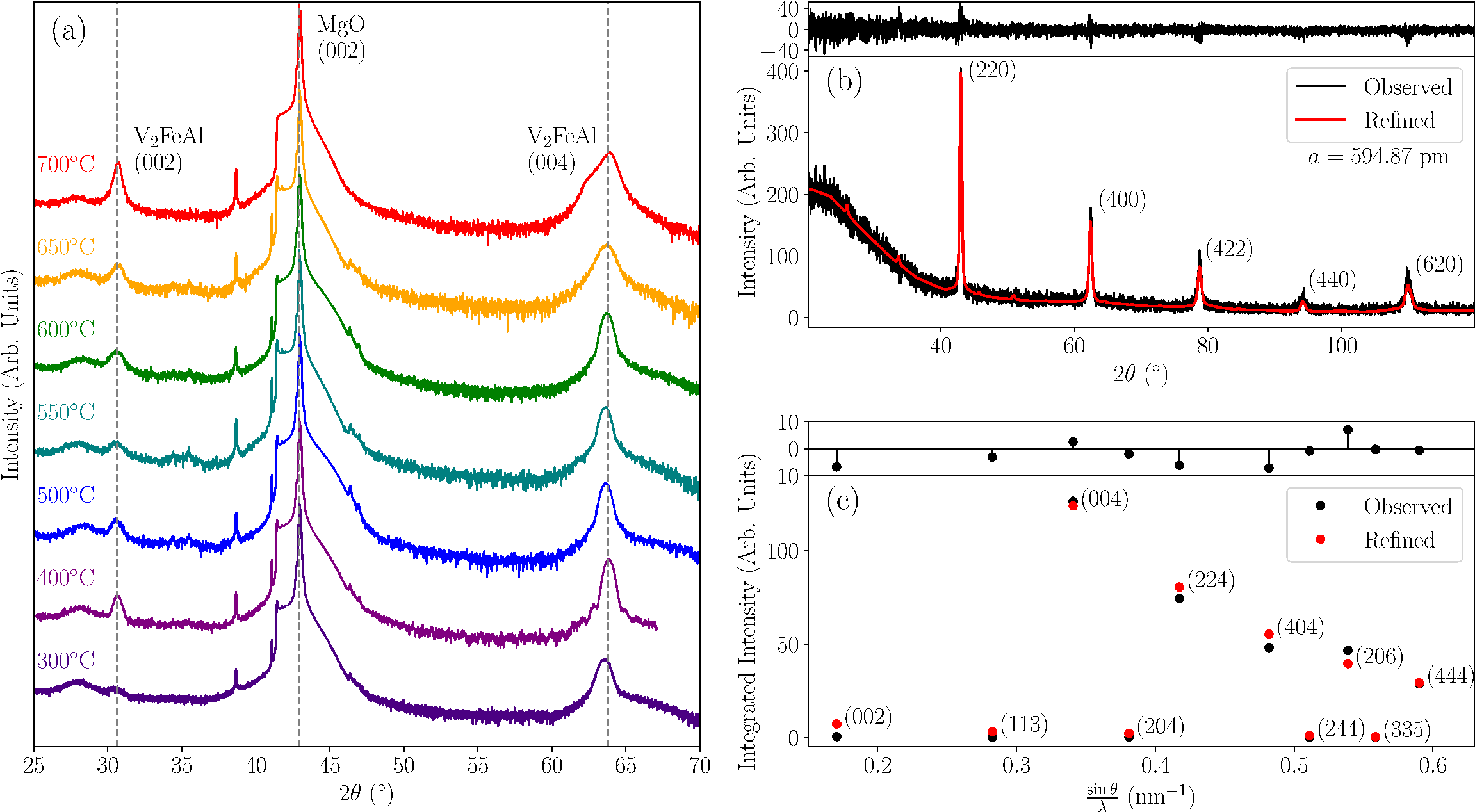}
	\caption{(a) X-ray diffraction patterns of V\textsubscript{2}FeAl thin-films grown at various deposition temperatures, (b) Rietveld refinement of V\textsubscript{2}FeAl powder diffraction pattern, (c) Rietveld refinement of several integrated peak intensities captured using grazing-incidence X-ray diffraction on V\textsubscript{2}FeAl thin-films}\label{fig:V2FeAl_XRD}
\end{figure*}

Figure~\ref{fig:V2FeAl_XRD} (a) shows X-ray diffraction patterns for thin-film V\textsubscript{2}FeAl as a function of deposition temperature. 
The best quality films are obtained at a temperature of \SI{400}{\degreeCelsius}, with this sample showing Laue oscillations about the {(004)} reflection indicating a large out-of-plane crystalline coherence length, the out-of-plane lattice parameter for this sample was found to be \SI{583.36}{\pico\meter}.
All subsequent data presented for ordered V\textsubscript{2}FeAl thin-films is measured on this sample.
Reciprocal space mapping of the V\textsubscript{2}FeAl {(204)} and {($\bar{2}\bar{2}4$)} reflections confirm an epitaxial relationship with the MgO{(001)} single-crystal substrate (i.e.\ a = b = \SI{595.55}{\pico\meter} $\approx\sqrt{2}\cdot$a\textsubscript{MgO}). 
The thin-films therefore show a large tetragonal distrotion induced by the substrate with c/a = 0.9795. 
As can clearly be seen for the sample deposited at \SI{700}{\degreeCelsius}, phase separation is beginning to occur with the {(004)} reflection showing a clear shoulder peak. 
Based upon the measured {(002)}/{(004)} ratio, the binary compound FeAl was determined to be the phase responsible for this peak.
Figure~\ref{fig:V2FeAl_XRD} (b) shows the powder diffraction pattern of a bulk V\textsubscript{2}FeAl sample, Rietveld analysis confirms that it forms the fully-disordered A2 structure (as indicated by the absence of a {(002)} reflection) with a cubic lattice parameter of \SI{594.87}{\pico\meter}, which corresponds well with the cubic lattice parameter predicted for the ferrimagnetic configuration of XA-type structured V\textsubscript{2}FeAl.
A number of thin-film samples deposited at lower temperatures also showed an unusually high {(002)}/{(004)} ratio, indicating the presence of FeAl, which can likely be attributed to poor epitaxial growth of V\textsubscript{2}FeAl. 
These samples also contain V\textsubscript{2}FeAl in the fully-disordered A2 form which we observe for bulk powders.
The disordered sample used for subsequent measurements was deposited at \SI{400}{\degreeCelsius}.

$\omega - 2\theta$ X-ray diffraction measurements were performed on an ordered V\textsubscript{2}FeAl thin-film to determine the integrated intensities of ten independent reflections.
After applying appropriate corrections to account for the illuminated sample area, the Lorentz polarization correction, and geometrical factors, the integrated peak intensities can be compared to those calculated from their structure factor.
Rietveld refinement can be performed to determine which structure type best matches the sample as shown in Figure~\ref{fig:V2FeAl_XRD} (c).
It was determined that the thin-film sample was closer to XA-type structure than L2\textsubscript{1} type structure, with the best fit lying somewhere between XA-type and the partially-disordered B2 structure.
However, owing to the similar X-ray scattering cross-sections of vanadium and iron, there are no significant differences in intensity between reflections from the two structures.
It is therefore impossible to determine the crystal structure decisively by X-ray diffraction using a laboratory source alone. 
X-ray reflectivity measurements found the films to have thickness of approximately \SI{15}{\nano\meter}, the ordered sample being \SI{14.70}{\nano\meter} thick and the disordered being \SI{15.78}{\nano\meter} thick. 
The films were all found to be smooth with RMS roughness values less than \SI{0.5}{\nano\meter}, and the thin-film densities were found to be around \SI{5.80}{\gram\per\centi\meter\cubed} (assuming full site occupancy, the nominal density is \SI{5.83}{\gram\per\centi\meter\cubed}).

\begin{figure*}[!ht]
	\centering
	\includegraphics[width=\textwidth]{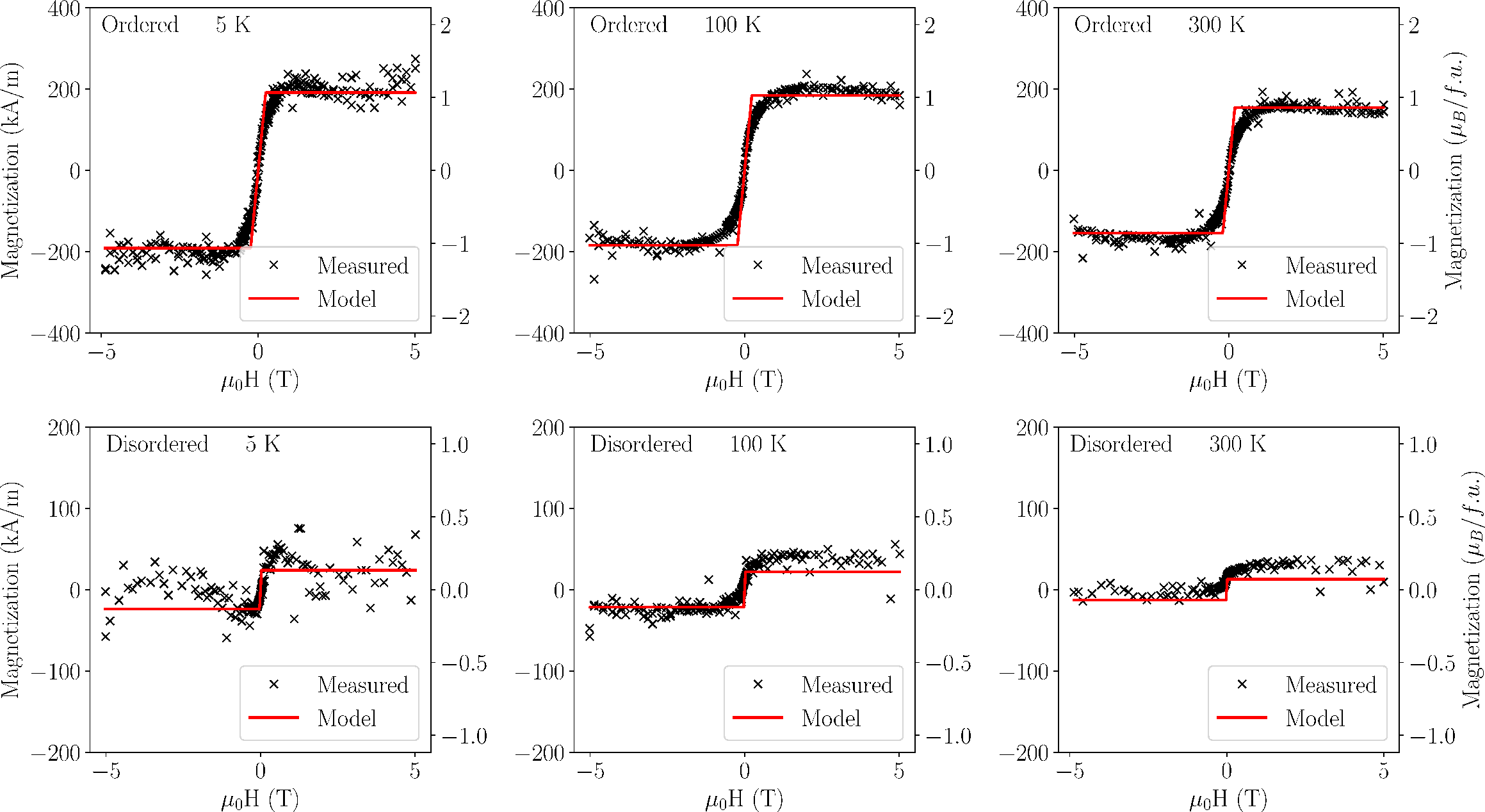}
	\caption{Out-of-plane magnetization loops with the ferromagnetic component of the fitting model for ordered and disordered thin-films of V\textsubscript{2}FeAl at \SI{5}{\kelvin}, \SI{100}{\kelvin}, and \SI{300}{\kelvin}.}\label{fig:V2FeAl_SQUID}
\end{figure*}

The magnetic properties of both bulk and thin-film samples of V\textsubscript{2}FeAl were measured using SQUID magnetometry. 
The measured magnetic moments of the thin-films were then corrected as described in the experimental section, and the anhysteretic spontaneous magnetization curves are presented in Figure~\ref{fig:V2FeAl_SQUID}.
The ordered thin-film shows a magnetization of approximately 1.0 \textmu\textsubscript{B} per formula unit at \SI{5}{\kelvin}, which falls gradually to approximately 0.9 \textmu\textsubscript{B} per formula unit at \SI{300}{\kelvin}, indicating a Curie temperature well in excess of room temperature (T\textsubscript{C} $\sim$ \SI{400}{\kelvin}). 
The disordered thin-film shows a magnetization of approximately 0.1 \textmu\textsubscript{B} per formula unit at \SI{5}{\kelvin}, about 10\% of the ordered films magnetization, with the magnetization falling to nearly zero at room temperature (T\textsubscript{C} $\sim$ \SI{330}{\kelvin}).
The bulk sample was found to be a Pauli paramagnet with a volume susceptibility of \textchi\textsubscript{v} = \SI{-2.95e-04}{}.

\begin{figure*}[!ht]
	\centering
	\includegraphics[width=\textwidth]{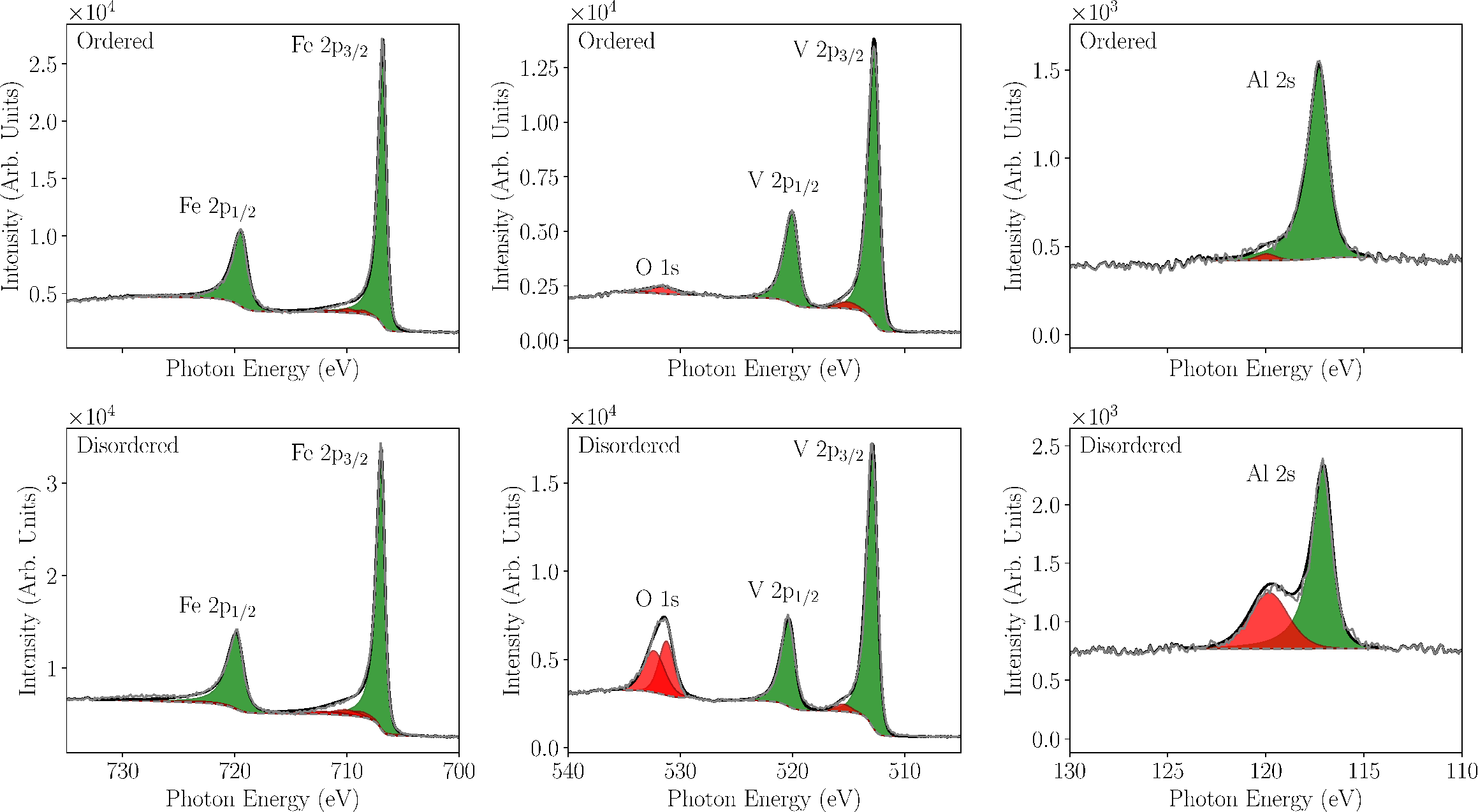}
	\caption{XPS spectra of the Fe 2p (left), V 2p and O 1s (centre), and Al 2s peaks (right) measured on both ordered (top row) and disordered (bottom row) V\textsubscript{2}FeAl thin films. The peaks shaded in green are metallic in nature, whereas those in red originate from oxygen or oxide species.}\label{fig:V2FeAl_XPS}
\end{figure*}

The XPS spectra of both ordered and disordered thin-films of V\textsubscript{2}FeAl are presented in Figure~\ref{fig:V2FeAl_XPS}.
Comparing the integrated areas of the Fe 2p, V 2p, and Al 2s peaks and applying their respective relative sensitivity factors, we find that the ordered sample contains slightly less vanadium than predicted based upon sample densities measured by XRR, but was still within 20\% of the nominal composition V\textsubscript{2}FeAl.
The disordered sample was found to contain approximately 17\% oxygen, significantly more than the ordered sample which contains approximately 3\% oxygen.
No significant contaminants other than oxygen were observed for either sample.
The vanadium, iron, and aluminium peaks all appear to be mostly metallic in nature, indicating both samples are electrically conductive.

\begin{figure*}[!ht]
	\centering
	\includegraphics[width=\textwidth]{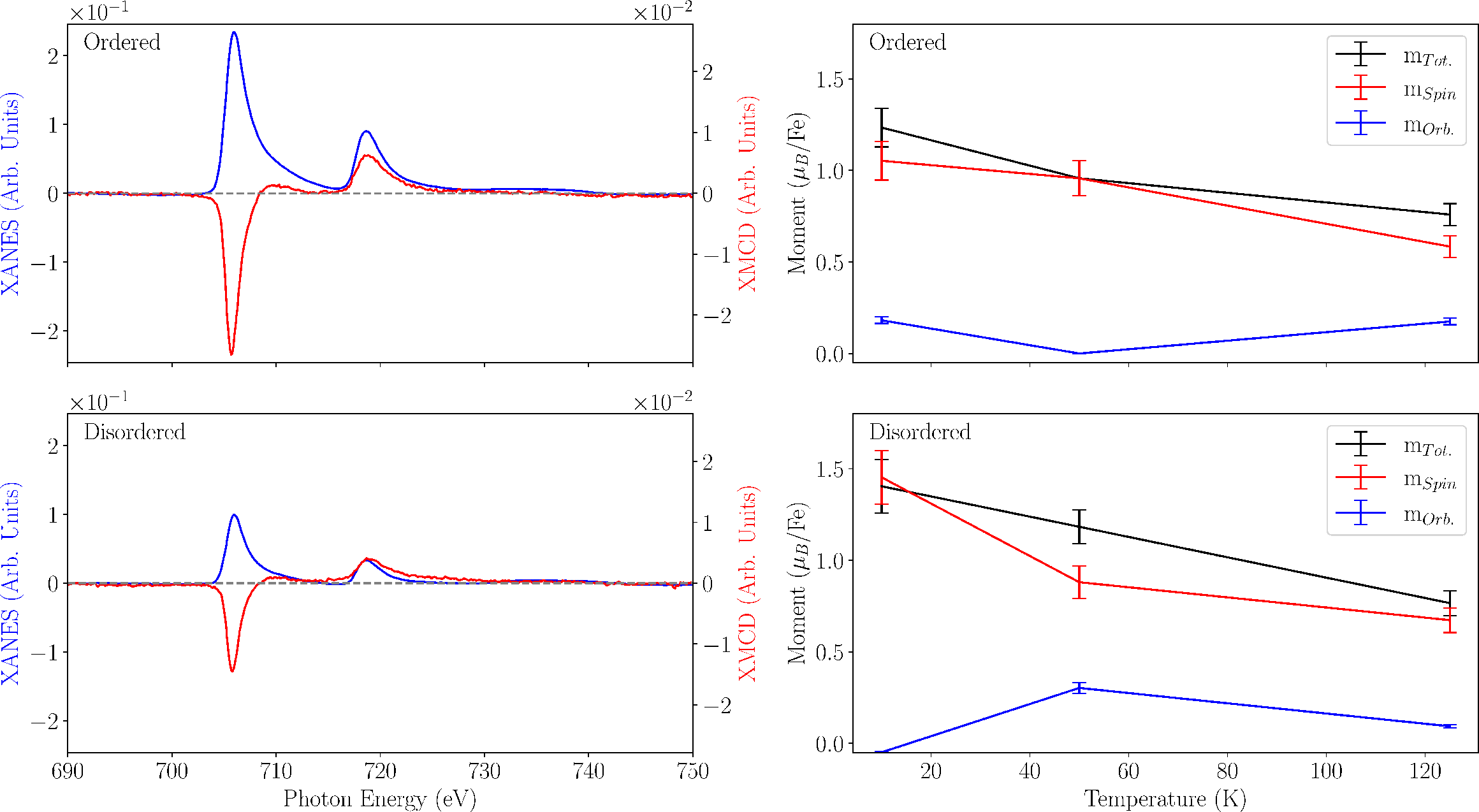}
	\caption{The X-ray absorption spectrum (with the background removed) and the dichroic XMCD signal measured on an ordered and disordered thin-film of V\textsubscript{2}FeAl (left), and the spin, orbital and total magnetic moments calculated using the sum rules as a function of temperature (right).}\label{fig:V2FeAl_XMCD}
\end{figure*}

Iron L-edge X-ray absorption spectroscopy and X-ray magnetic circular dichroism are shown in Figure~\ref{fig:V2FeAl_XMCD}. 
The panel on the left shows the XANES signal in blue (with the background signal removed), and the XMCD signal in red.
The panel on the right shows the spin and orbital moments (and their sum) obtained using the sum rules. 
The total iron moment for the ordered film at \SI{10}{\kelvin} is found to be 1.24 \textmu\textsubscript{B} per iron atom, which is in good agreement with the predicted moment for the ordered XA-type structure of 1.26 \textmu\textsubscript{B}.
Comparing this to the total moment measured by SQUID of 1.0 \textmu\textsubscript{B} per formula unit at \SI{5}{\kelvin}, we infer an average moment of -0.12 \textmu\textsubscript{B} per vanadium atom, which is somewhat lower than predicted for vanadium in the $4b$ and $4c$ sites {(-0.29 \textmu\textsubscript{B})} and much lower than the 1.88 \textmu\textsubscript{B} calculated for the $4d$ site.
At a temperature of \SI{125}{\kelvin}, the iron moment is reduced to 0.77 \textmu\textsubscript{B}, whereas the total moment measured by SQUID is approximately 0.9 \textmu\textsubscript{B} at \SI{100}{\kelvin}.
Conversely, the disordered sample shows an iron moment of 1.14 \textmu\textsubscript{B} per atom at \SI{10}{\kelvin}.
Given the small total moment of 0.1 \textmu\textsubscript{B} at \SI{5}{\kelvin}, the inferred moment on vanadium is -0.65 \textmu\textsubscript{B} which decreases with temperature to -0.33 \textmu\textsubscript{B} as the iron moment drops to 0.77 \textmu\textsubscript{B} at \SI{125}{\kelvin}.

\begin{figure*}[!ht]
	\centering
	\includegraphics[width=\textwidth]{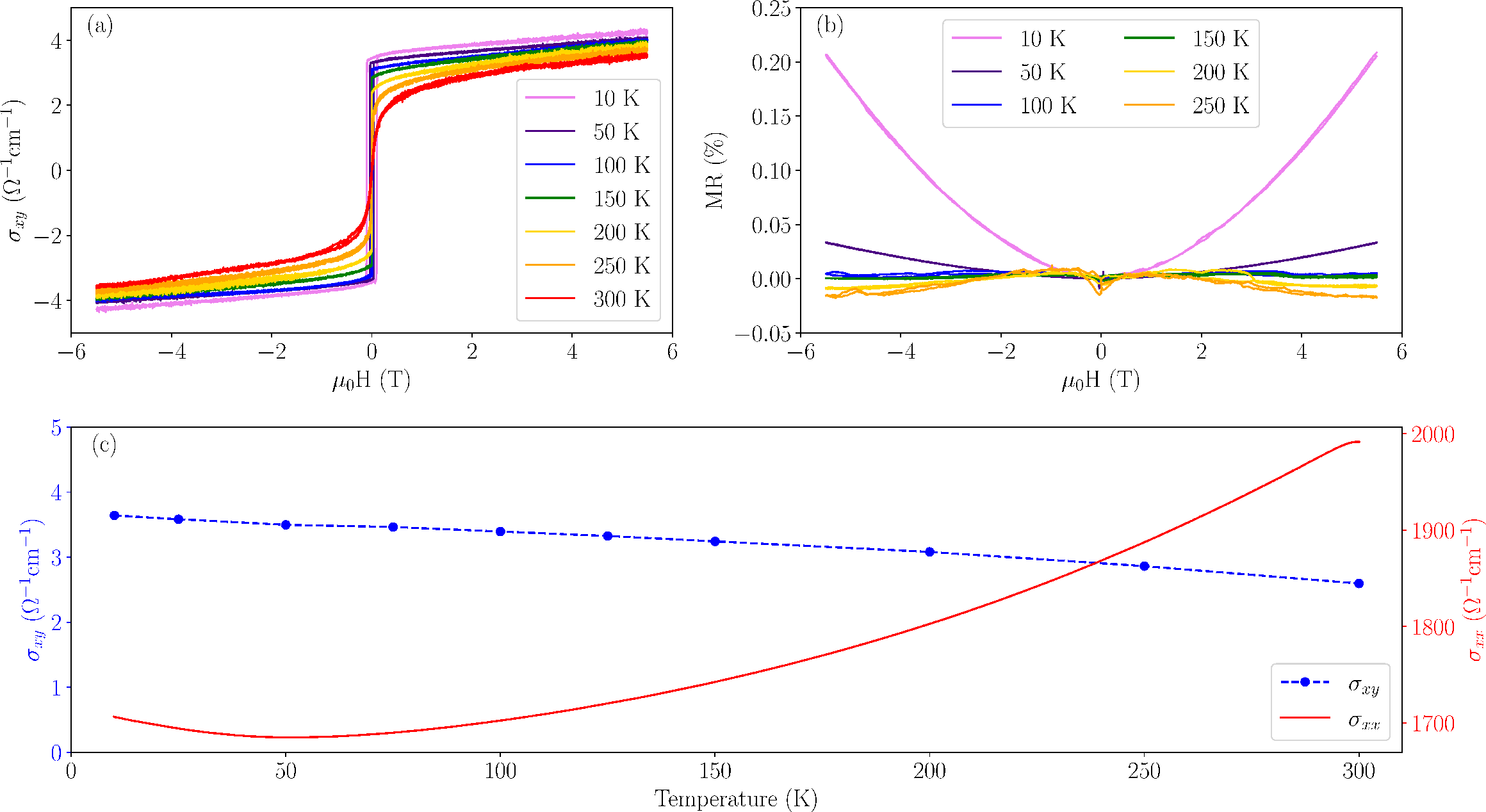}
	\caption{(a) Hall conductivity field loops at various temperatures, (b) magnetoresistance at various temperatures, (c) Hall conductivity (blue) and longitudinal conductivity (red) as a function of temperature measured in an applied field of \SI{5.5}{\tesla}.}\label{fig:V2FeAl_transport}
\end{figure*}

The magnetotransport properties of an ordered V\textsubscript{2}FeAl thin-film patterned into a Hall bar are shown in Figure~\ref{fig:V2FeAl_transport}.
The longitudinal conductivity is of order \SI{1e3}{\per\ohm\per\centi\meter}, indicating that V\textsubscript{2}FeAl is a bad metal, with the anomalous Hall effect being primarily due to impurity and defect scattering.
The temperature dependence of the anomalous Hall conductivity follows the magnetization, falling gradually from \SIrange{3.64}{2.59}{\per\ohm\per\centi\meter} in the temperature range \SIrange{10}{300}{\kelvin}.

\section{Discussion}

When grown in bulk and annealed at T\textsubscript{a} = \SI{650}{\degreeCelsius}, V\textsubscript{2}FeAl forms in the disordered cubic A2 structure. 
This is expected given the low enthalpy of formation of both ordered forms predicted by electronic structure calculations, and the entropy of disorder of $R\ln\Omega$ per mole (R is the universal gas constant, $\Omega$ the number of possible configurations of the system, in this case $\Omega=12$), which reduces the free energy by $RT_{a}\ln\Omega$ = \SI{19}{\kilo\joule\per\mole} at the annealing temperature of \SI{923}{\kelvin}.
Contrast this to the calculated total energy difference per atom shown in Figure~\ref{fig:V2FeAl_DFT}, which shows an energy difference of 159 meV between the ferrimagnetic L2\textsubscript{1} and XA-type structures corresponding to  a value of approximately \SI{15}{\kilo\joule\per\mole}.
The energy associated with disordering the system is therefore of the same order of magnitude as the energy difference between the different structures, indicating that formation of fully-ordered alloys is unlikely.

Bulk V\textsubscript{2}FeAl may be regarded as a high-entropy alloy like CrVTiAl quaternaries that are also Pauli paramagnets~\cite{Zhang2021}.
However, when deposited in thin-film form on an appropriate substrate, in our case MgO{(001)}, the induced tetragonal distortion stabilises a more highly ordered form of V\textsubscript{2}FeAl, allowing for ferrimagnetic ordering between the iron and vanadium atoms, as opposed to the paramagnetic behaviour observed in bulk samples.
These ordered thin-film samples show a structure which lies somewhere between the fully-ordered XA-type and the partially ordered B2 structure.
In all cases, XA-type structure appears to be preferred to L2\textsubscript{1} structure, which corroborates the predicted stability of the different configurations shown in Figure~\ref{fig:V2FeAl_DFT}.
X-ray photoelectron spectroscopy indicates that the disordered thin-films have a much higher concentration of surface oxygen than the ordered films.
It is unclear whether this oxygen is the result of contamination during the deposition process, or diffusion from poor quality MgO substrates.
The larger fraction of oxide species in the disordered sample means there will be less electrons available in the conduction band as they are bonding with the oxygen, and as a result a charge transfer between the iron and the other elements leaving more iron 3d core holes.
This explains the approximately 40\% more holes in the disordered sample which was determined from the magnitude of the XANES edge-jumps
The magnetic properties of the thin-film samples is highly dependent upon their ordering, with the ordered thin-film sample showing a saturation magnetization an order of magnitude greater than the disordered film at \SI{5}{\kelvin} as well as a higher Curie temperature.

It is apparent that disorder has a significant effect on the magnetic properties of V\textsubscript{2}FeAl.
The ordered thin-film at low temperature, which shows a total moment of 1.0 \textmu\textsubscript{B} per formula unit and an iron moment of 1.24 \textmu\textsubscript{B} per atom, from which we infer a vanadium moment of -0.12 \textmu\textsubscript{B} per atom.
Comparing this to the calculated moments presented in Table~\ref{tab:V2FeAl_DFT_Disorder}, we find the iron moment to be in good agreement with the predicted moment for a fully-ordered XA-type structure.
At the same time, there is no indication of a large vanadium moment, and the measured moments are also slightly lower than those predicted for vanadium in the $4b$ and $4c$ sites.
This is consistent with our assumption that DFT overestimates the vanadium moments in this system.
Conversely, the disordered thin-film sample shows a low-temperature total moment of 0.1 \textmu\textsubscript{B} per formula unit and an iron moment of 1.40 \textmu\textsubscript{B} per atom, from which we infer a vanadium moment of about -0.65 \textmu\textsubscript{B}.
The larger iron and vanadium moments and lower total moment relative to the ordered sample is consistent with what is observed in the DFT results as disorder is introduced to the system.
Comparison with Table~\ref{tab:V2FeAl_DFT_Disorder} suggests that the disordered sample is approximately 25\% XA-type structure.

\section{Conclusion}

Spontaneous magnetism in V\textsubscript{2}FeAl depends critically on establishing some ordering of the atomic constituents on the four face-centered cubic crystallographic sites of the Heusler structure.
Our electronic structure calculations show that the energy difference between the L2\textsubscript{1} and XA-type ordered structure is approximately \SI{15}{\kilo\joule\per\mole}.
Annealed bulk material with a high-entropy, fully-disordered A2 structure is found experimentally and it is a Pauli paramagnet.
Tetragonally-distorted thin-films grown on MgO substrates exhibit spontaneous magnetism that depends on the degree of disorder and oxidation.
Substrate templated thin-film samples with tetragonal distortion can stabilise a partly XA-ordered V\textsubscript{2}FeAl structure which shows a spontaneous magnetization with moments on both iron and vanadium.
The measured iron moments match well with DFT predictions, but the total moment is lower than predicted for a fully-ordered XA-type structure.
Based upon comparison to the binary FeV system with similar vanadium concentrations, we accredit this discrepancy to an overestimation of the moments of vanadium in the $4d$ sites when they are not fully occupied by vanadium.
The formation of the metastable ordered structure appears to be hindered by the presence of oxygen in the thin-film.
We believe that in reality the $4d$ vanadium moments are closer to those predicted for the $4b$ and $4c$ sites, resulting in a lower total moment.
A drawback of electronic structure calculations based on small supersctructures is that to capture the nuances of disorder in imperfectly-ordered alloys they would need to consider much larger unit cells than that of the basic Heusler structure.

\section{Acknowledgements}
The authors acknowledge financial support from Science Foundation Ireland through contract 16/IA/4534 ZEMS.
The authors acknowledge the support of Dr. Alevtina Smekhova of Helmholtz-Zentrum Berlin for performing X-ray absorption and dichroism measurements.

\bibliography{bibliography.bib}

\end{document}